\documentclass[a4paper,fleqn,usenatbib]{mnras}

\usepackage{mathptmx}

\usepackage[T1]{fontenc}
\usepackage{ae,aecompl}

\usepackage{graphicx}	
\usepackage{amsmath}	
\usepackage{amssymb}	

\title[Spectroscopy and detection of $c$-C$_2$H$_3$DO]{Rotational spectroscopy 
       of mono-deuterated oxirane ($c$-C$_2$H$_3$DO) and its detection towards IRAS 16293$-$2422~B}

\author[H. S. P. M{\"u}ller et al.]{
Holger S. P. M{\"u}ller,$^{1}$\thanks{E-mail: hspm@ph1.uni-koeln.de (HSPM)}
Jes K. J{\o}rgensen,$^{2}$ 
Jean-Claude Guillemin,$^{3}$ 
Frank Lewen$^{1}$ 
\newauthor{and Stephan Schlemmer$^{1}$}
\\
$^{1}$I.~Physikalisches Institut, Universit{\"a}t zu K{\"o}ln,
      Z{\"u}lpicher Str. 77, 50937 K{\"o}ln, Germany\\
$^{2}$Niels Bohr Institute, University of Copenhagen,\\ {\O}ster Voldgade 5$-$7, 1350 Copenhagen K, Denmark\\
$^{3}$Univ Rennes, Ecole Nationale Sup{\'e}rieure de Chimie de Rennes, CNRS, ISCR$-$UMR 6226, 35000 Rennes, France
}

\date{Accepted XXX. Received YYY; in original form ZZZ}

\pubyear{2022}

\begin{document}
\label{firstpage}
\pagerange{\pageref{firstpage}--\pageref{lastpage}}
\maketitle

\begin{abstract}
  We prepared a sample of mono-deuterated oxirane and studied its   rotational spectrum 
  in the laboratory between 490~GHz and 1060~GHz in order to improve its spectroscopic 
  parameters and consequently the calculated rest frequencies of its rotational transitions. 
  The updated rest frequencies were employed to detect $c$-C$_2$H$_3$DO for the first time 
  in the interstellar medium in the Atacama Large Millimetre/submillimetre Array (ALMA) 
  Protostellar Interferometric Line Survey (PILS) of the Class~0 protostellar system 
  IRAS 16293$-$242. Fits of the detected lines using the rotation diagrams yield 
  a temperature of $T_{\rm rot} = 103 \pm 19$~K, which in turn agrees well with 125~K
  derived for the $c$-C$_2$H$_4$O main isotopologue previously. The $c$-C$_2$H$_3$DO to 
  $c$-C$_2$H$_4$O ratio is found to be $\sim$0.15 corresponding to a D-to-H ratio 
  of $\sim$0.036 per H atom which is slightly higher than the D-to-H ratio of species 
  such as methanol, formaldehyde, ketene and but lower than those of the larger complex 
  organic species such as ethanol, methylformate and glycolaldehyde. This may reflect that 
  oxirane is formed fairly early in the evolution of the prestellar cores. 
  The identification of doubly deuterated oxirane isotopomers in the PILS data may be 
  possible judged by the amount of mono-deuterated oxirane and the observed trend that 
  multiply deuterated isotopologues have higher deuteration rates than their 
  mono-deuterated variants. 
\end{abstract}

\begin{keywords}
astrochemistry -- ISM: molecules -- line: identification -- 
methods: laboratory: molecular -- 
ISM: abundances -- ISM: individual objects: IRAS 16293$-$2422. 
\end{keywords}



\section{Introduction}
\label{intro}

The cyclic molecule oxirane, $c$-C$_2$H$_4$O, has been detected in several places in 
warm regions associated with low- and high-mass star formation. It was detected first towards 
the molecule-rich Galactic centre source Sagittarius B2(N) \citep{det-c-C2H4O_1997}. 
It was observed subsequently in the warm gas associated with several high-mass star-forming 
regions \citep{more_obs_c-C2H4O_1998,still_more_obs_c-C2H4O_2001}. 
\citet{with-O_Miguel_2008} reported its identification in three rotationally cold, but
kinetically warm Galactic centre sources. 
\citet{PILS_COMs_2017} found the molecule towards the low-mass protostellar source 
IRAS 16293$-$2422 more recently and \citet{c-C2H4O_L1689B_2019} reported its detection 
in a prestellar core.

The report by \citet{PILS_COMs_2017} was one of the early results of the Protostellar 
Interferometric Line Survey (PILS): PILS is an unbiased molecular line survey around 
345~GHz carried out with the Atacama Large Millimetre/submillimetre Array (ALMA) 
to study the physical conditions and the molecular complexity of the Class~0 
solar-type protostellar system and astrochemical template source IRAS 16293$-$2422 
\citep{PILS_overview_2016}. Through the high sensitivity of PILS and narrow lines 
towards one component of IRAS~16293-2422 (limiting line confusion) a number of species 
have been detected for the first time, including the organohalogen compound methyl chloride 
(CH$_3$Cl) \citep{PILS_MeCl_2017}, nitrous acid (HONO) \citep{PILS_HONO_2019} and, 
tentatively, 3-hydroxypropenal \citep{3-hydroxypropenal_det_2022}.

An important contribution of PILS has also been the numerous isotopologues of 
organic molecules, in particular deuterated variants, that were detcted for 
the first time in the interstellar medium (ISM). Deuterated molecules are particularly 
noteworthy among the isotopic species because the enrichment of deuterium in dense 
molecular clouds has attracted considerable interest for many years as the degree 
of deuteration has been viewed as an evolutionary tracer in low-mass star-forming 
regions and/or a means to trace the formation histories of complex organic molecules 
\citep[e.g.,][]{deuteration_1989,deuteration_2005,deuteration_2007,taquet14,deuteration_and_c-C3H2_2018}.

Deuterated isotopologues are particularly prominent towards low-mass star 
forming regions due to the low temperatures promoting deuterium fractionation. 
With its rich spectrum IRAS 16293$-$2422 is therefore one of the prime targets 
to search for deuterated molecules: and already through single-dish observations 
dedicated studies led to the first detections of methanol species such as CHD$_2$OH 
\citep{det_CHD2OH_2002}, CD$_3$OH \citep{det_CD3OH_2004}, mono-deutered dimethyl 
ether, CH$_2$DOCH$_3$, \citep{CH3OCHD2_rot_det_2013}, the ground state terahertz 
transitions of H$_2$D$^+$ \citep{H2D+_THz_2014} and HD$_2^+$ \citep{HD2+_THz_2017} 
and NHD and ND$_2$ \citep{NHD_ND2_det_2020}.

The deuterated species detected in the PILS data are the mono-deuterated isotopomers 
of the oxygen-bearing organics glycolaldehyde \citep{PILS_overview_2016}, ethanol, 
ketene, formic acid and of mono-deuterated acetaldehyde species CH$_3$CDO 
\citep{PILS_div-isos_2018} and CH$_2$DCHO \citep{det_CH2DCHO_2019,manigand20}, 
of the nitrogen-bearing organics isocyanic acid DNCO and the mono-deuterated 
isotopomers of formamide \citep{PILS_formamide_2016} and the cyanamide isotopologue 
HDNCN \citep{PILS_cyanamide_2018} and sulfur-containing species such as the hydrogen 
sulfide isotopologue HD$^{34}$S \citep{PILS_sulfur_2018}. 
Also, the PILS data reveal the presence of doubly-deuterated organics including 
the methyl cyanide species CHD$_2$CN \citep{PILS_nitriles_2018}, the methyl 
formate species CHD$_2$OCHO \citep{PILS_dideu-MeFo_2019} and the dimethyl ether 
species CHD$_2$OCH$_3$ \citep{CH3OCHD2_rot_det_2021} and enable new and more 
accurate constraints on the doubly- and triply-deuterated variants of methanol 
in the warm gas close to the protostars \citep{CHD2OH_catalog_2022,CD3OH_rot_2022}. 
These systematic studies also enabled a more detailed comparison across the different 
species. It was seen for example that the degree of deuteration in a given molecule 
(referenced to one H atom) does not change for structurally different H atoms 
within uncertainties for several organics 
\citep[e.g.,][]{PILS_overview_2016,PILS_div-isos_2018}. It also appears that different 
types of organics can be categorised in groups according to their D/H ratios, which 
in turn may reflect their underlying formation mechanisms \citep{PILS_div-isos_2018}.

It appeared plausible to identify mono-deuterated oxirane ($c$-C$_2$H$_3$DO) in 
the PILS data judged by the strength of the $c$-C$_2$H$_4$O lines \citep{PILS_COMs_2017}. 
Very limited transition frequencies, however, were available only until fairly recently 
\citep{c-C2H4O_div-isos_rot_1974}; only 20 transitions were published up to 60~GHz with 
$J \le 8$ and of these were only three strong $R$-branch transitions with $J \le 2$. 
\citet{c-C2H3DO_rot_2019} reported on a laboratory spectroscopic investigation of 
$c$-C$_2$H$_3$DO which provided us with the means to search for this isotopic species. 
While the data were extensive and accurate enough to identify \mbox{$c$-}C$_2$H$_3$DO 
with certainty in the PILS data we were faced with a small, but non-negligible deviation 
of the velocity offset in comparison to those of $c$-C$_2$H$_4$O and other molecules. 
We decided to synthesize a sample of $c$-C$_2$H$_3$DO to investigate its 
rotational spectrum further in order to improve the spectroscopic parameters and 
hoping to be able to resolve the deviation.

The remainder of the article is laid out as follows. Section~\ref{lab-spec} 
deals with our laboratory spectroscopic investigations, Section~\ref{astro-obs} 
describes the astronomical search for $c$-C$_2$H$_4$O and Section~\ref{conclusion} 
summarises the main findings of the paper.


\begin{figure}
\begin{center}
	\includegraphics[width=.65\columnwidth]{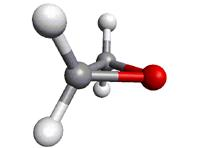}
    \caption{Model of the $c$-C$_2$H$_4$O molecule. The C atoms are shown in grey, 
             The H atoms in light grey and the O atom in red. The O atom and the mid-point 
             of the CC bond are on the $b$-axis.}
    \label{fig:molecule}
\end{center}
\end{figure}


\section{Laboratory investigations}
\label{lab-spec}

\subsection{Spectroscopy of oxirane}
\label{spec-details}


Oxirane ($c$-C$_2$H$_4$O) is also known as epoxyethane, ethylene oxide, dimethylene 
oxide and oxacyclopropane. Its rotational spectrum was studied quite extensively 
\citep{c-C2H4O_rot_1974,c-C2H4O_rot_isos_1974,c-C2H4O_FASSST_1998,c-C2H4O_FIR_2012,c-C2H4O_rot_2022}. 
The molecule has $C_{\rm 2v}$ symmetry with its dipole moment of 1.90~D along the $b$-axis 
\citep{c-C2H4O_S_rot_dip_1951}. This value includes a small upward correction of 0.02~D 
\citep{c-C2H4O_rot_2022} caused by the difference of the OCS reference value used in 
\citet{c-C2H4O_S_rot_dip_1951} compared to more modern values \citep{OCS_dip_1985,OCS_dip_1986}. 
The main isotopologue is an asymmetric top rotor of the oblate type with 
$\kappa = (2B - A - C)/(A - C) = +0.4093$ which is quite far from the oblate symmetric limit 
of +1. The four equivalent H nuclei lead to \textit{ortho} and \textit{para} spin-statistics 
with intensity weight ratios of $5 : 3$. The \textit{ortho} and \textit{para} levels 
are described by $K_a + K_c$ being even and odd, respectively.

Substitution of one of the four H atoms by D lowers the symmetry to $C_1$ and leads to a 
chiral species, similar to the related molecule methyl oxirane, also known as propylene oxide 
etc., recently detected as the first chiral molecule in space \citep{det_Me-Oxirane_2016}. 
We derived from the structural parameters of the molecule that the rotation of the principal 
inertial axis system of $c$-C$_2$H$_3$DO with respect to that of $c$-C$_2$H$_4$O leads to a 
small $\mu _a$ dipole moment component of 0.36~D, a very small $\mu _c = 0.08$~D and an 
almost unchanged $\mu _b = 1.86$~D. Mono-deuterated oxirane is also an asymmetric top rotor 
of the oblate type with $\kappa = +0.2042$ even more distant from the oblate symmetric limit 
of +1. No non-trivial spin-statistics occur in the rotational spectrum of $c$-C$_2$H$_3$DO 
because of the low symmetry of the isotopologue.


\subsection{Synthesis of mono-deuterated oxirane}
\label{synthesis}


Oxirane-1-$d$ was synthesized in a two-step reaction: the synthesis of 2-chloroethanol-1-$d$ 
and the dehydrochlorination of the latter to oxirane-1-$d$. We imagined an original approach for 
the second step in order to be able to synthesize small quantities of the desired product.

\textbf{2-Chloroethanol-1-$d$}. The chloroacetaldehyde in water (50\%) (10 g, 64~mmol) was 
extracted with diethyl ether (50~mL) and the organic phase was dried over MgSO$_4$. 
Sodium borodeuteride (1.0~g, 24~mmol) and dry diethyl ether (10~mL) were introduced under 
nitrogen into a three-necked flask and the chloroacetaldehyde in diethyl ether was added 
dropwise (bubbling was observed). The reaction mixture was refluxed for 3~h, then hydrolysed 
with 3~N-H$_2$SO$_4$, made alkaline with aqueous sodium carbonate, extracted with diethyl ether 
and dried over MgSO$_4$. The solvent was carefully removed under vacuum and the purification 
was carried out on a vacuum line (0.1~mbar) with a trap cooled to $-$60$^{\rm o}$C to selectively 
trap the 2-chloroethanol-1-$d$. Two distillations gave 1.8~g pure product with an isotopic purity 
higher than 98\%. The yield is 90\% based on sodium borodeuteride.

$^1$H NMR (CDCl$_3$, 400~MHz) $\delta$ 3.09 (s, brd, 1H, OH), 3.66 (d, 2H, $^3$J$_{\rm HH} = 5.2$~Hz, 
CH$_2$Cl), 3.85 (dt, 1H, $^3$J$_{\rm HH} = 5.2$~Hz, $^2$J$_{\rm HD} = 1.9$~Hz, CHD). 
$^{13}$C NMR (CDCl$_3$, 100~MHz) $\delta$ 46.6 ($^1$J$_{\rm CH} = 175.5$~Hz (t), CH$_2$Cl), 
62.6 ($^1$J$_{\rm CD} = 22.0$~Hz (t), $^1$J$_{\rm CH} = 143.7$~Hz (d), CHD).

\textbf{Oxirane-1-$d$}. 2-Chloroethanol-1-$d$ (1.63~g, 20~mmol) was then vaporised in a vacuum line (0.1~mbar) 
onto t-BuOK in powder heated to 90$^{\rm o}$C \citep[for similar experiments see][]{4synth_2019}. 
A trap cooled to $-$100$^{\rm o}$C removed selectively the impurities (mainly t-butanol) and oxirane-1-$d$ 
was selectively condensed in a trap cooled at 77~K at a yield of 0.60~g or 67\%.

$^1$H NMR (CDCl$_3$, 400~MHz) $\delta$ 2.54 (m, 3H, CH$_2$O and CHD). $^{13}$C NMR (CDCl$_3$, 100~MHz) 
$\delta$ 40.2 ($^1$J$_{\rm CD} = 26.9$~Hz (t), $^1$J$_{\rm CH} = 175.5$~Hz (d), CHD), 
40.4 ($^1$J$_{\rm CH} = 175.5$~Hz (t), CH$_2$).


\subsection{Rotational spectroscopy}
\label{exptl}


The measurements were carried out at room temperature in a 5~m long single path Pyrex glass cell 
equipped with high-density polyethylene windows. Two frequency multipliers (VDI Inc.) driven by a 
Rhode \& Schwarz SMF~100A microwave synthesizer were used as sources in the region 490$-$1060~GHz. 
A closed cycle liquid He-cooled InSb bolometer (QMC Instruments Ltd) was employed as a detector. 
Additional information on this spectrometer system is available in \citet{CH3SH_rot_2012}.

The entire region from 490~GHz to 750~GHz was covered at a total pressure of 1~Pa and with low 
bolometer sensitivity to record mainly medium strong lines of $c$-C$_2$H$_3$DO. Strong lines that 
displayed (potentially) weak indications of saturation in these spectral recordings were rerecorded 
individually at a pressure of 0.2~Pa. Very strong lines with clear signs of saturation were 
remeasured at a pressure of 0.05~Pa. The same region was covered again at a high bolometer 
sensitivity and a pressure of 4~Pa for an improved signal-to-noise ratios (S/N) of weak lines.

Very weak lines in this region with relatively large uncertainties after inclusion of the stronger 
lines were recorded individually at a pressure of 4~Pa and with longer integration times. Similar 
measurements were finally carried out between 810~GHz and 1060~GHz.

The S/N of our spectral recordings is very high such that the uncertainties assigned to the 
transition frequencies are mostly dominated by the line shape. Very symmetric and isolated 
lines have uncertainties of 5~kHz, a value that appeared even conservative for several of the 
strong and very strong lines. We have shown in studies of H$_2$CO \citep{H2CO_rot_2017} and 
H$_2$CS \citep{H2CS_rot_2019} that uncertainties of 10~kHz can be achieved quite regularly 
at higher frequencies of 1.3$-$1.5~THz. Baseline effects or other lines in the vicinity 
affect the line shape. The assigned uncertainties of slightly less symmetric lines were 
10$-$20 kHz while 30~kHz or 50~kHz were employed for some lines with more pronounced 
asymmetries.


\subsection{Treatment of previous data}
\label{prev-data}


Fitting and calculation of $c$-C$_2$H$_3$DO transition frequencies was performed with Pickett's 
SPFIT and SPCAT programmes \citep{spfit_1991}. We refitted the previous $c$-C$_2$H$_3$DO data 
\citep{c-C2H4O_div-isos_rot_1974,c-C2H3DO_rot_2019} in order to derive transition frequencies, 
their uncertainties and intensities for our present laboratory studies. 
We assumed 50~kHz for the data from \citet{c-C2H4O_div-isos_rot_1974} in analogy to the experimental 
lines of the main isotopic species \citep{c-C2H4O_rot_1974} from the same author and based on our 
judgement \citep{c-C2H4O_rot_2022} of the $c$-C$^{13}$CH$_4$O and $c$-C$_2$H$_4$$^{18}$O data from 
\citet{c-C2H4O_div-isos_rot_1974}. Two transition frequencies from that work with large residuals of 
0.20~MHz and 0.72~MHz were omitted. We employed the millimetre wave data from \citet{c-C2H3DO_rot_2019} 
as provided by one of the authors of that study because the data published as supplementary data to 
\citet{c-C2H3DO_rot_2019} were in units of wavenumbers and truncated at the $10^{-5}$~cm$^{-1}$ or 
300~kHz level. These data were estimated to have uncertainties of about 200~kHz in the original work. 
A fit of these data alone, however, had an rms of 30~kHz \citep{c-C2H3DO_rot_2019}. We carried out 
several trial fits with different uncertainties assigned to these data and different sets of 
spectroscopic parameters which suggested that the uncertainties of these millimetre wave data are 
more like 10~kHz possibly even better. We used 10~kHz as uncertainties initially and 5~kHz later.

The majority of the lines from \citet{c-C2H3DO_rot_2019} were obtained in the far-infrared (FIR) 
region by Fourier transform spectroscopy (FTS). The authors estimated these data to be accurate 
to 0.00010~cm$^{-1}$ or about 3~MHz. Their fit as well as our trial fits yielded rms values 
of slightly better than 0.00008~cm$^{-1}$ ($\approx 2.4$~MHz); we employed therefore 
0.00008~cm$^{-1}$ as uncertainties of the FIR data. All but two of the FIR transition 
frequencies refer to unresolved asymmetry doublets, but only one of the two transitions was given 
in the line list. This is fine for asymmetry splittings well within the frequency uncertainties 
but introduces an error otherwise. We assigned two transitions to all unresolved asymmetry doublets 
which essentially doubled the FIR line list. There is essentially no effect on the parameter 
uncertainties because if two or more lines are blended to one they are treated as one piece 
of information in SPFIT. The effect on the rms of the FIR data turned out to be modest because 
the frequencies of most unresolved asymmetry doublets were identical within the quoted digits.

We need to point out that FTS requires calibration of the spectra in order to achieve reliable 
transition frequencies. The FIR data in \citet{c-C2H3DO_rot_2019} were calibrated with lines 
of residual water in the optical path, a common procedure if enough suitable water lines are 
present. These are usually compared to reference data to determine a calibration factor. 
The rms of the FIR transitions deteriorated in the early stages of adding lines from our 
measurements to the fit until a value of 0.00016~cm$^{-1}$ was reached and an average shift 
of the FIR data of 0.00015~cm$^{-1}$ or 4.5~MHz lower than the calculated frequencies was 
obtained which is indicative of a small but non-negligible calibration error. Trial fits 
with the aim to minimize the average shift yielded a recalibration factor of 1.0000037. 
The inverse of this value can be entered as infrared calibration factor in the parameter 
files of SPFIT. The rms of the FIR lines returned to the initial value of slightly better 
than 0.00008~cm$^{-1}$ after this correction.

We employed the scaled spectroscopic parameters up to sixth order from a quantum-chemical 
calculation \citep{c-C2H2DO_calc_param_2014} as starting values as was done by 
\citet{c-C2H3DO_rot_2019}. It turned out that rotational and quartic centrifugal 
distortion parameters plus the sextic parameter $\Phi _{KJ}$ were needed to minimize 
the rms error as measure for the quality of the fit. Some other sextic distortion 
parameters appeared to be determined, meaning their uncertainties were a sufficiently 
small fraction of their values but had only a small impact on the rms error. 
Releasing different sets of sextic distortion parameters resulted in transition 
frequencies that differed by around 0.2~MHz for the lines we had identified in our 
astronomical data, significant at line widths of about 1~MHz. We decided to revisit 
the rotational spectrum of \mbox{$c$-}C$_2$H$_3$DO exactly because of these deviations.


\begin{figure}
	\includegraphics[width=.95\columnwidth]{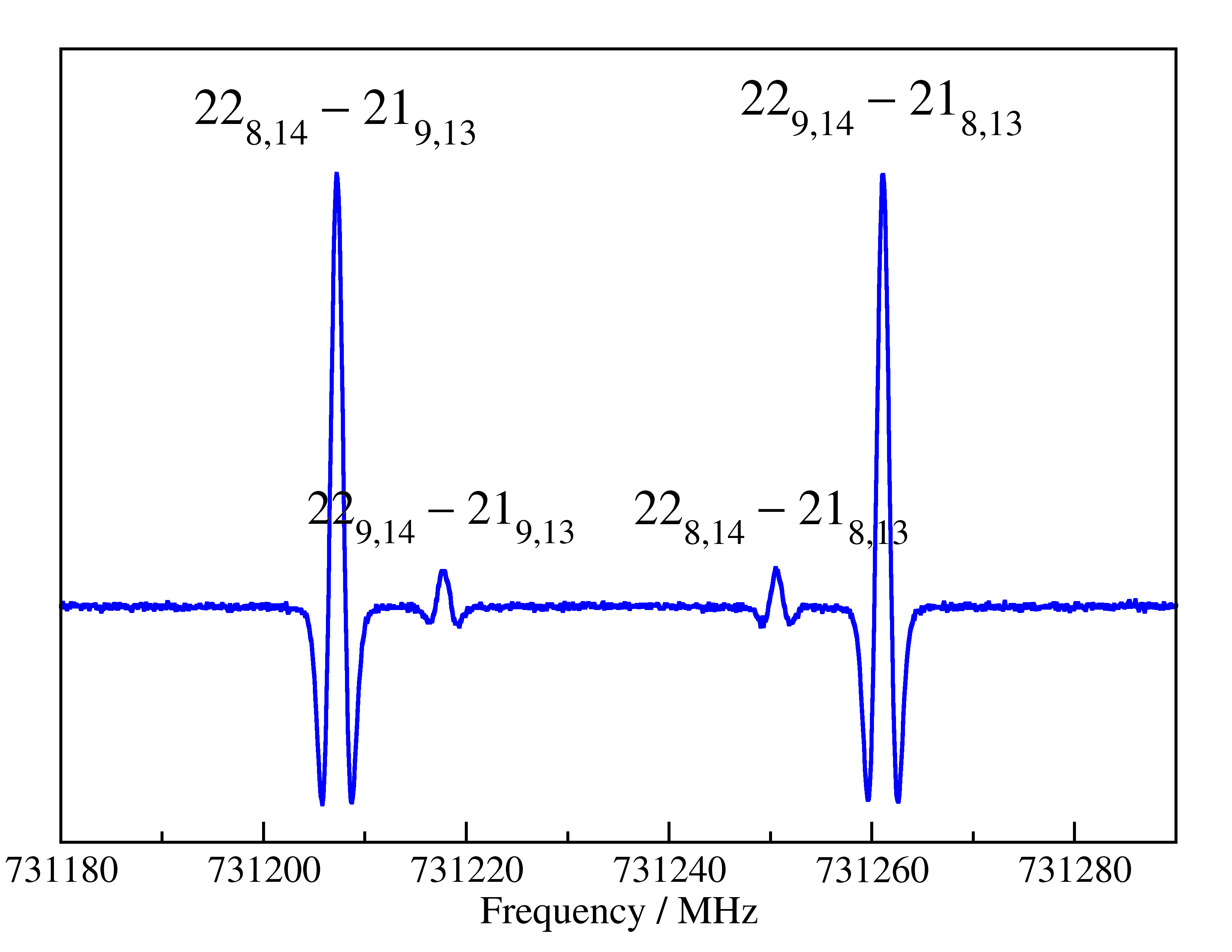}
    \caption{Section of the rotational spectrum of $c$-C$_2$H$_3$DO. Two stronger $b$-type 
             transitions approach oblate pairing and are accompanied by much weaker $a$-type 
             transitions between them.}
    \label{fig:oblate-pairing}
\end{figure}


\subsection{Observed spectrum and determination of spectroscopic parameters}
\label{lab-results}


The rotational spectrum of $c$-C$_2$H$_3$DO is sparse on the level of the strong and very strong 
lines. These are all $b$-type $R$-branch ($\Delta J = +1$) transitions in the 490$-$750~GHz region. 
The $b$-type transitions have $\Delta K_a = \pm1$, $\pm3$ etc. and $\Delta K_c = \pm1$, $\pm3$ 
etc. More specifically, these transitions are mostly prolate paired (unresolved asymmetry doublets 
both having the same upper and lower $J$ and $K_a$ values) with $J''$ about 10 to 15 and low values 
of $K_c$ or oblate paired transitions (unresolved asymmetry doublets both having the same upper 
and lower $J$ and $K_c$ values) with $J''$ about 16 to 27 and low values of $K_a$. 
These transitions were found close to their calculated positions even if they were shifted 
from their calculated positions by a few megahertz. The spectroscopic parameters were improved 
after each round of assignments leading to improved calculations of the rotational spectrum.

The $b$-type transitions approaching oblate pairing, these are transitions close but distinct 
in frequency having the same $J$ and $K_c$ values, are accompanied by much weaker \mbox{$a$-}type 
transitions between them, as is shown in Fig.~\ref{fig:oblate-pairing}. These patterns are easy 
to recognize and may be used for assignment purposes if the calculated spectrum is uncertain. 
The relatively strong $a$-type transitions have $\Delta K_a = 0$ and $\Delta J = \Delta K_c = +1$. 
We also assigned several much weaker $a$-type transitions with $\Delta K_a = +2$ and 
$\Delta J = -\Delta K_c = +1$. Among the weaker $b$-type transitions were $Q$-branch transitions 
($\Delta J = 0$), $R$- and $Q$-branch transitions with $\Delta K_a = -\Delta K_c = \pm3$, 
high-$K_a$ transitions with $\Delta J = -\Delta K_a = +1$ and with other selection rules.


\begin{figure}
	\includegraphics[width=.95\columnwidth]{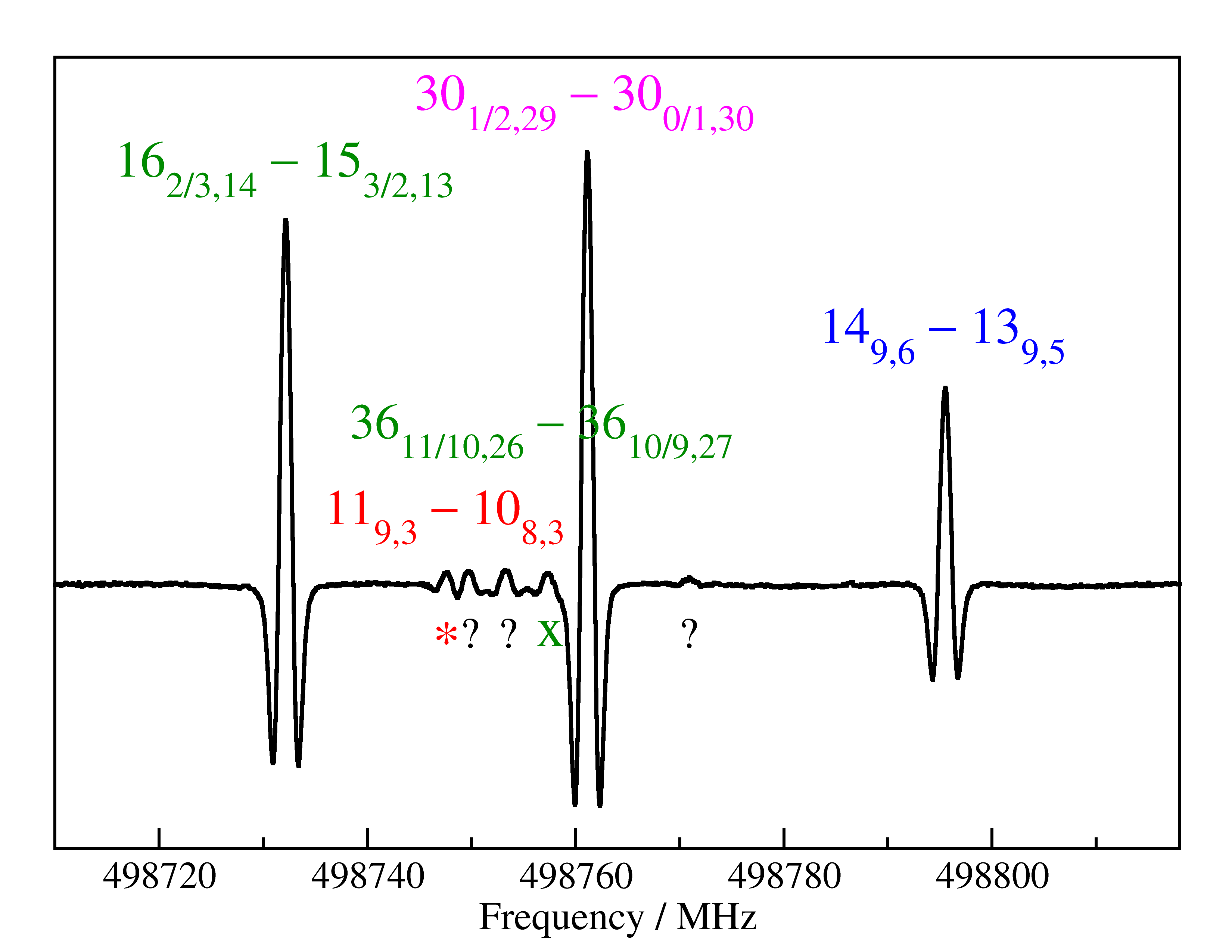}
    \caption{Section of the rotational spectrum of $c$-C$_2$H$_3$DO. Two oblate paried $b$-type 
             transitions are shown in the centre. A weaker $a$-type transition is visible to the right 
             and a very weak $c$-type transition marked with an asterisk to the left. The relative 
             intensities support the magnitudes of the dipole moment components calculated from the 
             structure. Two lines (quantum numbers in green, the weak one labelled with an x) were 
             assigned to $c$-C$_2$H$_4$O at a level of $\sim$1.5\% of the sample pressure. 
             Clearly visible but unassigned lines are labelled with question marks.}
    \label{fig:dipole_etc}
\end{figure}


Eventually we assigned a fair fraction of the strongest though still weak or very weak $c$-type 
transitions with $\Delta J = \Delta K_a = +1$ and $\Delta K_c = 0$. In the case of transitions 
approaching prolate pairing, the $c$-type transitions are between the close-lying much stronger 
$b$-type transitions. Fig.~\ref{fig:dipole_etc} displays a $c$-type transition of $c$-C$_2$H$_3$DO
accompanied by an $a$-type transition and two oblate paired $b$-type transitions supporting 
the magnitudes of the dipole moment components calculated from the structure. 
Also shown are two lines of $c$-C$_2$H$_4$O; we conclude from their intensities that about 
1.5\% of the sample pressure is caused by the fully hydrogenated isotopologue. 
We included afterwards the very weak transitions recorded in the 490$-$750~GHz region and 
subsequently those from the 810$-$1060~GHz region into the fit.


\begin{table*}
\begin{center}
\caption{Number of transitions (No. T), number of distinct frequencies (No. F), maximum quantum 
         numbers, rms (kHz, 10$^{-3}$~cm$^{-1}$)$^a$ and rms error (unitless) of the individual 
         data sets and of the global fit employing the III$^l$ representation in the S reduction.}
\label{tab-quality}
\begin{tabular}[t]{lcccccr@{}lc}
\hline 
source$^b$                           & No. T & No. F & $J_{\rm max}$ & $K_{a,\rm max}$ & $K_{c,\rm max}$ & \multicolumn{2}{c}{rms} & rms error \\
\hline 
MW \citep{c-C2H4O_div-isos_rot_1974} &   18  &   18  &        8      &          5      &          4      &          46&.2          &   0.924   \\
mmW \citep{c-C2H3DO_rot_2019}        &  112  &  112  &       23      &         14      &         10      &           5&.02         &   1.004   \\
FIR \citep{c-C2H3DO_rot_2019}        &  782  &  392  &       57      &         40      &         56      &           0&.070        &   0.879   \\
sub-mmW, present                     & 1504  & 1091  &       58      &         43      &         55      &           9&.13         &   0.884   \\
global                               & 2416  & 1613  &       58      &         43      &         56      &            &$c$         &   0.892   \\
\hline
\end{tabular}\\[2pt]
\end{center} 
$^a$ Units of kHz for all but the FIR data.\\
$^b$ The abbreviations stand for microwave, millimetre wave, far-infrared and submillimetre wave respectively.\\
$^c$ The rms values are 10.40~kHz and 0.00007~cm$^{-1}$ ($\approx$ 2.1~MHz) for MW to sub-mmW data and FIR data respectively.
\end{table*}


\begin{table*}
\begin{center}
\caption{Present experimental spectroscopic parameters (MHz) of mono-deuterated oxirane and previous values 
         from an experimental study$^a$ (prev. exptl.) and from a quantum-chemical study after scaling$^b$ (scaled ai).}
\label{tab-spec-param}
\renewcommand{\arraystretch}{1.10}
\begin{tabular}[t]{lr@{}lcr@{}lr@{}lr@{}ll}
\hline 
 \multicolumn{3}{c}{S reduction, III$^l$} & & \multicolumn{7}{c}{A reduction, I$^r$} \\
\cline{1-3} \cline{5-11}
Parameter & \multicolumn{2}{c}{present} & & \multicolumn{2}{c}{present} & 
\multicolumn{2}{c}{prev. exptl.} & \multicolumn{2}{c}{scaled ai} & Parameter \\
\hline
$A$                      &  24252&.670221~(39) & &   24252&.647815~(44) &   24252&.64765$^c$  &  24250&.3320 & $A$                       \\
$B$                      &  19905&.493836~(40) & &   19905&.523291~(40) &   19905&.52126$^c$  &  19906&.4347 & $B$                       \\
$C$                      &  13327&.592078~(53) & &   13327&.583649~(53) &   13327&.58134$^c$  &  13327&.2181 & $C$                       \\
$D_K \times 10^3$        &     18&.028074~(55) & &      25&.46266~(19)  &      25&.478~(21)   &     17&.458  & $\Delta_K \times 10^3$    \\
$D_{JK} \times 10^3$     &  $-$52&.12685~(11)  & &      17&.25673~(12)  &      17&.252~(22)   &     22&.401  & $\Delta_{JK} \times 10^3$ \\
$D_J \times 10^3$        &     41&.656664~(99) & &      17&.04199~(11)  &      17&.0744~(69)  &     15&.689  & $\Delta_J \times 10^6$    \\
$d_1 \times 10^3$        &      8&.308953~(27) & &      13&.231691~(56) &      13&.250~(10)   &     16&.214  & $\delta_K \times 10^3$    \\
$d_2 \times 10^3$        &   $-$0&.743758~(8)  & &       4&.742162~(17) &       4&.7399~(21)  &      4&.124  & $\delta_J \times 10^3$    \\
$H_K \times 10^9$        & $-$114&.912~(41)    & &     808&.09~(31)     &     870&.~(60)      &    757&.5    & $\Phi_K \times 10^9$      \\
$H_{KJ} \times 10^9$     &    220&.075~(113)   & &  $-$835&.25~(36)     &  $-$960&.~(60)      & $-$868&.8    & $\Phi_{KJ} \times 10^9$   \\
$H_{JK} \times 10^9$     & $-$123&.995~(124)   & &     141&.97~(16)     &     186&.~(21)      &    170&.3    & $\Phi_{JK} \times 10^9$   \\
$H_J \times 10^{9}$      &     21&.586~(95)    & &      12&.364~(105)   &      14&.1~(21)     &      6&.0    & $\Phi_J \times 10^{9}$    \\
$h_1 \times 10^{9}$      &  $-$10&.980~(26)    & &      30&.210~(103)   &      61&.7$^d$      &     61&.7    & $\phi_K \times 10^9$      \\
$h_2 \times 10^{9}$      &     26&.227~(13)    & &      62&.667~(69)    &      81&.1$^d$      &     81&.1    & $\phi_{JK} \times 10^9$   \\
$h_3 \times 10^{9}$      &  $-$15&.071~(3)     & &       4&.797~(14)    &       2&.2$^d$      &      2&.2    & $\phi_J \times 10^{9}$    \\
$L_K \times 10^{12}$     &   $-$1&.578~(11)    & &    $-$9&.13~(25)     &        &            &      &       & $L_K \times 10^{12}$      \\
$L_{KKJ} \times 10^{12}$ &      4&.280~(37)    & &       6&.51~(27)     &        &            &      &       & $L_{KKJ} \times 10^{12}$  \\
$L_{JK} \times 10^{12}$  &   $-$4&.424~(53)    & &       1&.55~(20)     &        &            &      &       & $L_{JK} \times 10^{12}$   \\
$L_{JJK} \times 10^{12}$ &      2&.546~(39)    & &    $-$2&.10~(6)      &        &            &      &       & $L_{JJK} \times 10^{12}$  \\
$L_J \times 10^{12}$     &   $-$0&.922~(27)    & &    $-$0&.170~(30)    &        &            &      &       & $L_J \times 10^{12}$      \\
$l_1 \times 10^{15}$     &    399&.2~(74)      & &    $-$4&.571~(60)    &        &            &      &       & $l_K \times 10^{12}$      \\
$l_2 \times 10^{15}$     & $-$209&.4~(47)      & &       2&.243~(41)    &        &            &      &       & $l_{KJ} \times 10^{12}$   \\
$l_3 \times 10^{15}$     &    177&.6~(18)      & &    $-$1&.191~(25)    &        &            &      &       & $l_{JK} \times 10^{12}$   \\
$l_4 \times 10^{15}$     &  $-$26&.7~(3)       & &   $-$34&.7~(36)      &        &            &      &       & $l_J \times 10^{15}$      \\
                         &       &             & &       1&.305~(103)   &        &            &      &       & $P_K \times 10^{15}$      \\
                         &       &             & &    $-$1&.949~(171)   &        &            &      &       & $P_{KKJ} \times 10^{15}$  \\
                         &       &             & &       0&.851~(99)    &        &            &      &       & $P_{KJ} \times 10^{15}$   \\
$p_5 \times 10^{18}$     &   $-$1&.95~(13)     & &        &             &        &            &      &       &                           \\
\hline
\end{tabular}\\[2pt]
\end{center} 
Numbers in parentheses are one standard deviation in units of the least significant figure.\\
$^a$ \citet{c-C2H3DO_rot_2019}; fit of all data.\\
$^b$ \citet{c-C2H2DO_calc_param_2014}.\\
$^c$ Rotational parameters kept fixed to values from a fit to microwave and millimetre wave data.\\
$^d$ Off-diagonal sextic distortion parameters kept fixed to scaled values from \citet{c-C2H2DO_calc_param_2014}.
\end{table*}


The strategy to determine spectroscopic parameters has been the usual for the most part. We search 
for the spectroscopic parameter whose release, if the value was kept fixed to an estimated value, 
or whose inclusion reduces the rms error of the fit the most in order to keep the set of parameters 
as small and as unique as possible. This strategy works mostly very well in cases of asymmetric 
rotors close to the prolate limit of $-$1. It works sometimes less well for asymmetric rotors of 
the oblate type presumably because spectroscopic parameters of a given order are similar in magnitude 
and correlated. The extreme of an alternative approach is to include all reasonable spectroscopic 
parameters or the remaining part thereof into the fit and omit subsequently those parameters 
which display large uncertainties or whose omission from the parameter set increases the rms error 
only by small amounts. This latter approach was employed at least at one stage of the fittings.

Several spectroscopists view Watson's A reduction in an oblate representation as the natural 
choice to fit an asymmetric rotor of the oblate type. Several studies on molecules such as 
H$_2$S \citep{H2S_FIR_1994}, dimethylsulfoxide \citep{DMSO_2010}, the lowest energy conformer 
of 2-cyanobutane \citep{2-CAB_rot_2017} or the main isotopologue of oxirane \citep{c-C2H4O_rot_2022} 
revealed that this choice of reduction and representation is frequently a rather poor one. 
All other choices yielded better fits or required fewer parameters. In the case of oxirane all 
other common combinations of reduction and representation, the S reduction in the oblate III$^l$ 
representation and the A and S reductions in the prolate I$^r$ representation yielded satisfactory 
fits albeit the A reduction in I$^r$ required two parameters less than the two combinations 
involving S reductions. The A reduction in I$^r$ was in fact the preferred choice in several 
studies on $c$-C$_2$H$_4$O \citep{c-C2H4O_FIR_2012,c-C2H4O_rot_2022} and $c$-C$_2$H$_3$DO 
\citep{c-C2H2DO_calc_param_2014,c-C2H3DO_rot_2019}. We used the scaled parameters from a 
quantum-chemical calculation \citep{c-C2H2DO_calc_param_2014} as starting values in our 
present study on $c$-C$_2$H$_3$DO exactly for this reason. We also tried out the S reduction 
in the oblate III$^l$ representation to test a more natural choice of representations. 
Both choices resulted for a long time in fits of similar quality employing the same number 
of spectroscopic parameters. Later fits in I$^r$ and the A reduction required one and even 
two parameters less than fits of the same line lists in III$^l$ and the S reduction. 
The inclusion of our transition frequencies in the 810$-$1060~GHz altered the situation; 
now the III$^l$ fit in the S reduction required two parameters less than the I$^r$ fit 
in the A reduction.

We evaluated as final steps the impact of the previous data on the fit and the appropriateness 
of the assigned uncertainties. Omission of all previously reported transition frequencies 
changed the parameter values within the uncertainties and increased the parameter 
uncertainies only slightly; the largest increase occurred for the $C$ rotational parameter 
with slightly more than 10\%. The previous transition frequencies were retained in the final fit 
nevertheless because their impact was not negligible. Table~\ref{tab-quality} displays the extent 
and the upper quantum numbers of the individual data sets and of the fit as a whole. 
It also demonstrates that the data were fitted within the assigned uncertainties on average with 
a marginal exception for the millimetre wave data from \citet{c-C2H3DO_rot_2019}. 
However, the rms error of this data set exceeds 1.0 by such a small amount that an alteration 
of the uncertainties is not warranted. The quality of the A reduction fit in the I$^r$ 
representation is very similar; the rms error of the overall fit is 0.901.

The resulting spectroscopic parameters are gathered in Table~\ref{tab-spec-param} together 
with the previous experimental values from \citet{c-C2H3DO_rot_2019} and the scaled values 
from a quantum-chemical study \citep{c-C2H2DO_calc_param_2014}.

The fit files and a calculation of the rotational spectrum of $c$-C$_2$H$_3$DO have been 
deposited as supplementary material to this article together with an explanatory file. 
These files as well as additional files are also available in the Cologne Database for Molecular 
Spectroscopy (CDMS)\footnote{https://cdms.astro.uni-koeln.de/} \citep{CDMS_2005,CDMS_2016}.


\subsection{Discussion of the laboratory spectroscopic results}
\label{lab-discussion}


Our submillimetre spectroscopic data enabled us to determine full sets of very accurate 
spectroscopic parameters up to eighth order supplemented by one or three decic parameters 
respectively. The parameter set in the III$^l$ representation employing Watson's S reduction 
should be favoured over the set in the I$^r$ representation employing the A reduction because 
two fewer parameters are needed for a satisfactory fit. We point out that our experience in 
the course of the present study indicated that this choice may depend on details of the line 
list. The I$^r$ A reduction combination was the favoured one in fits of the main isotopologue 
\citep{c-C2H4O_rot_2022}. It is interesting to note that fits of the rotational spectrum of 
\mbox{$c$-}C$_2$H$_4$O required at most a full set of parameters up to eighth order 
(the disfavoured A reduction in the III$^l$ representation is an exception) whereas the present 
fits of $c$-C$_2$H$_3$DO required decic distortion parameters.

The present laboratory spectroscopic investigation improves the calculation of the rotational 
spectrum of \mbox{$c$-}C$_2$H$_3$DO considerably. The stronger transition frequencies are sufficiently 
accurate up to at least 1.5~THz which is sufficient for all astronomical observations dedicated 
to this isotopologue. The Boltzmann peak at 125~K is at $\sim$460~GHz and the strongest 
transitions are oblate paired transitions with $K_c = J$. The peak shifts to $\sim$960~GHz at 
room temperature and the strongest transitions are now prolate paired transitions with $K_a = J$.

The spectroscopic parameters in the I$^r$ representation employing Watson's A reduction can 
be compared with results from the recent experimental study on $c$-C$_2$H$_3$DO 
\citep{c-C2H3DO_rot_2019} and with results from a quantum-chemical calculation 
\citep{c-C2H2DO_calc_param_2014}. 
The comparison is favourable for the rotational and quartic centrifugal distortion parameters 
from the previous experimental study and reasonable for the sextic centrifugal distortion 
parameters both for the values kept fixed to scaled values from a quantum-chemical calculation 
and for those that were floated. The scaled values from \citet{c-C2H2DO_calc_param_2014} 
agree reasonably well with ours on average. The scaling improved the agreement for some 
parameters and deteriorated it for others. Our choice for the scaled values was motivated 
by the aspect that such scaling usually reduces deficiencies caused by the theoretical 
model or created by the neglect of vibrational effects in the common quantum-chemical 
calculations of centrifugal distortion parameters. Such scaling works usually well for heavy 
atom substitutions but not as well for H to D substitutions as may be seen in the example of 
the cyclopropenone isotopologues \citep{c-H2C3O_rot_2021}.

The frequencies calculated from the parameters in the S reduction and the III$^l$ representation 
for the strong transitions in the range of PILS (329.1 to 362.9~GHz; see Section~\ref{astro-obs}) 
differ little from our previous calculations based only on the data available through 
\citet{c-C2H3DO_rot_2019} for prolate paired transitions, in particular those with $K_a = J$. 
Deviations exceeding 300~kHz were observed for oblate paired transitions, in particular those 
with $K_c = J$. \citet{c-C2H3DO_rot_2019} provided also calculations of strong rotational transitions 
of $c$-C$_2$H$_3$DO between 300 and 400~GHz in their supplementary material. 
The frequencies of prolate paired transitions agree usually quite well with our current ones; 
but the deviations increase to nearly 70~MHz for the oblate paired transitions with 
$J = K_c = 13 - 12$. Such large deviations are difficult to explain in light of their FIR data 
combined with the fact that interpolation in quantum numbers is usually not a problem.


\begin{figure}
	\includegraphics[width=.92\columnwidth]{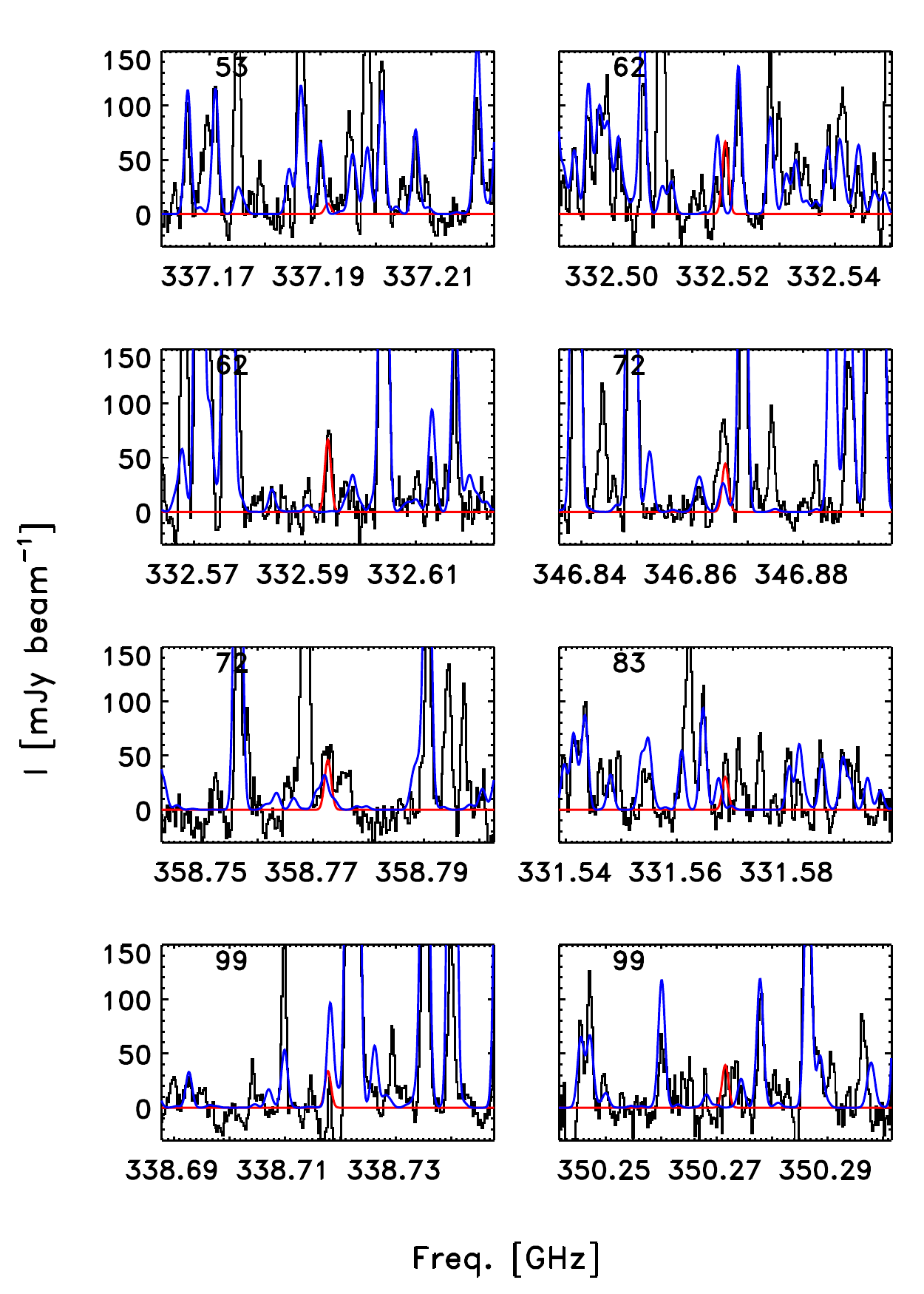}\\
	\includegraphics[width=.92\columnwidth]{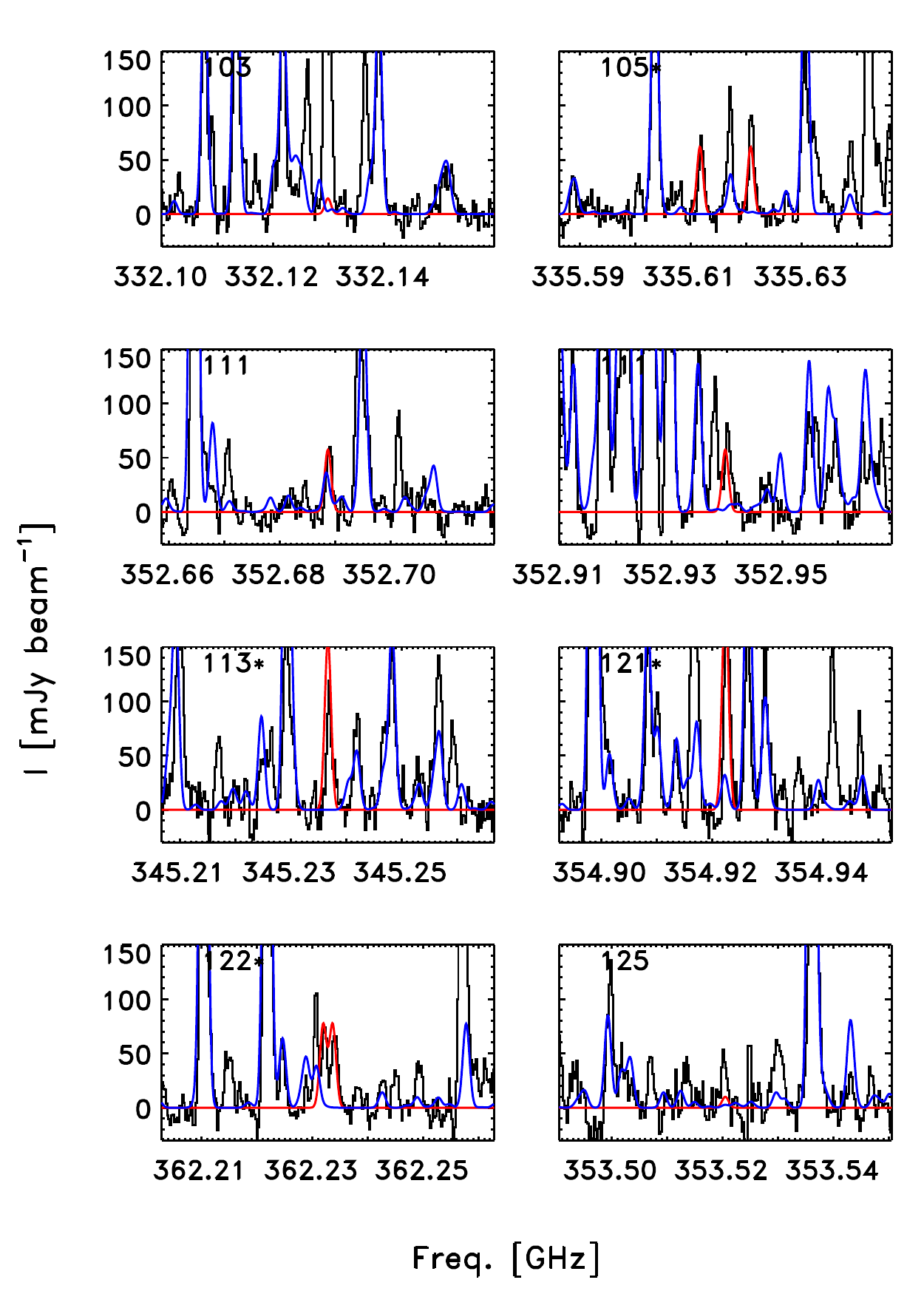}
    \caption{Sections of the Band~7 PILS data slightly offset from the continuum peak 
             displaying stronger lines of $c$-C$_2$H$_3$DO. The observed data are shown 
             in black, the modelled $c$-C$_2$H$_3$DO lines in red and the models of all 
             other assigned lines in blue.}
    \label{fig:astro-spec}
\end{figure}


\section{Astronomical search for mono-deuterated oxirane}
\label{astro-obs}

\subsection{Overview of the observations}
\label{details-obs}


We employed a calculation of the rotational spectrum of \mbox{$c$-}C$_2$H$_3$DO based 
on our results described in Section~\ref{lab-spec} to search for this isotopologue 
in data from PILS.  As described in introduction, PILS is an unbiased molecular
line survey of the protostellar object IRAS 16293$-$2422 carried out
with ALMA in its Cycle~2 (project id: 2013.1.00278.S, PI: J.~K.~J{\o}rgensen).  
The source is separated into two main components, the more prominent Source~A 
(in itself a binary source; \citealt{wootten89}) and the secondary 
Source~B, separated by about 5\arcsec\ (700~au), and both commonly 
viewed as Class~0 protostars. 
The survey covers part of Band~7 from 329.1 to 362.9~GHz at 0.2~km~s$^{-1}$ 
spectral and $\sim$0.5\arcsec \ angular resolution. An overview of the data 
and their reduction as well as first results from that survey are presented 
in \citet{PILS_overview_2016}. 
Also considered were Band~6 observations toward IRAS~16293-2422 in its
Cycle~4 (project id: 2016.1.01150.S, PI: V.~Taquet) targeting a few
selected windows at 232.71--234.18~GHz, 234.92--235.38~GHz,
235.91--236.38~GHz and 236.37--236.84~GHz, previously presented in
\citet{taquet18}. We refer to that paper for further details about
those data and their reduction. 

We focus on Source~B in the present work because its narrow lines ($\sim$1~km~s$^{-1}$) 
make it an ideal target for searches for rare species. The position that we have used almost
exclusively in investigations of Source~B is slightly offset from its continuum peak 
because opacity of the continuum is less severe, absorption lines are fewer and weaker 
and the lines are particularly narrow (see, e.g., 
\citealt{PILS_overview_2016,PILS_div-isos_2018,PILS_nitriles_2018}).


\subsection{Astronomical results}
\label{astro-results}

The search for $c$-C$_2$H$_3$DO towards IRAS 16293$-$2422~B was
carried out by fitting synthetic spectra to the PILS data similar to
other searches for complex molecules in this source.  The synthetic
spectra were calculated assuming the molecules are in local
thermodynamical equilibrium (LTE) which is sensible given the high
densities in the molecular cloud surrounding the protostar that is
probed by our ALMA observations \citep{PILS_overview_2016}.  We
assumed $T_{\rm rot} = 125$~K in our initial calculations as was
determined for $c$-C$_2$H$_4$O \citep{PILS_COMs_2017}. We employed a
linewidth (FWHM) of 1~km~s$^{-1}$ and a velocity offset relative to
the local standard of rest of 2.6~km~s$^{-1}$ for the same reasons.
Several of the stronger transitions of $c$-C$_2$H$_3$DO in the Band~7
data of PILS are shown in Fig.~\ref{fig:astro-spec} and are summarized
in Table~\ref{fig:astro-spec}.  A large fraction of these were found
to be unblended or only slightly blended. As indicated in
Section~\ref{intro} and Section~\ref{lab-discussion}, our own
calculation based on data available through \citet{c-C2H3DO_rot_2019}
had an accuracy that was sufficient to identify $c$-C$_2$H$_3$DO
towards IRAS 16293$-$2422~B but was somewhat limited to achieve a
satifactory analysis.

\begin{figure}
  \includegraphics[width=\columnwidth]{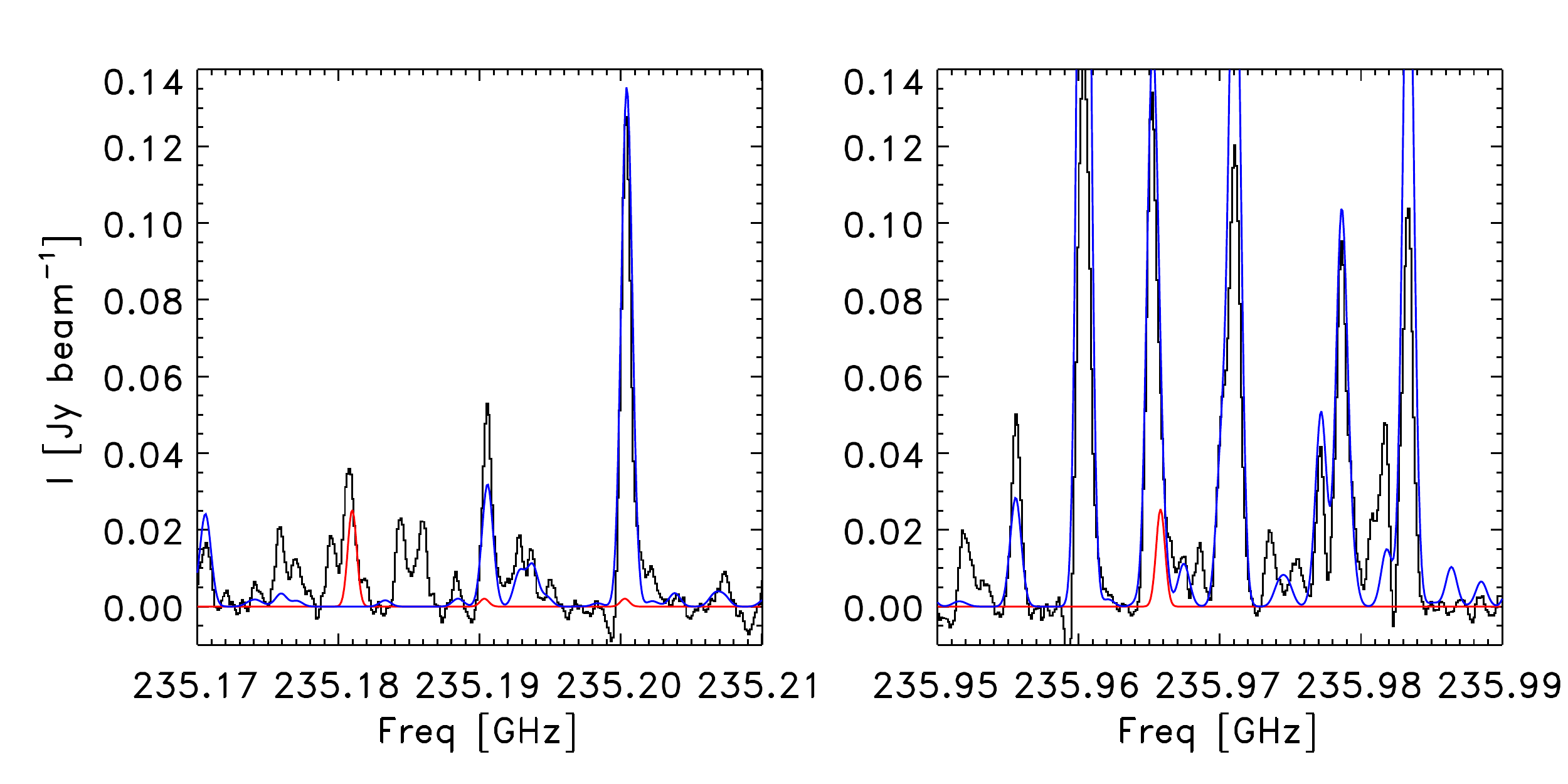}
  \caption{Comparison of the predicted spectrum for $c$-C$_2$H$_3$DO 
    (red line) to the observations at Band~6. The blue spectrum 
    represents line emission from the other species identified in the 
    PILS programs survey so-far. One transition at 235.18~GHz is 
    cleanly detected whereas others are at 235.96~GHz are blended with 
    a prominent line of methylformate.}\label{band6_figure}
\end{figure}

\begin{figure}
	\includegraphics[width=\columnwidth]{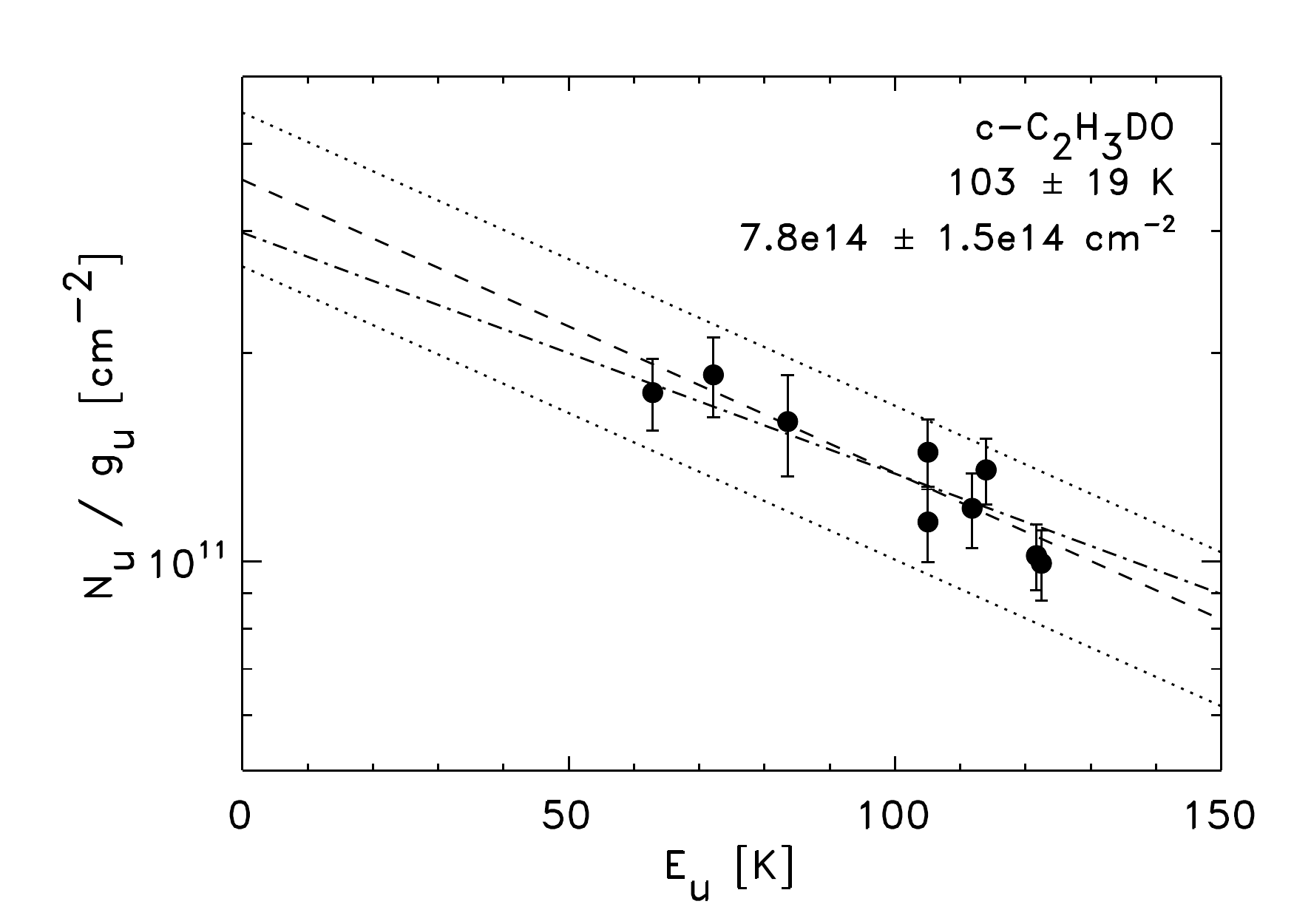}
    \caption{Rotation diagram for unblended or slightly blended lines of $c$-C$_2$H$_3$DO.}
    \label{fig:rotdia}
\end{figure}

The analysis yielded a $c$-C$_2$H$_3$DO column density 
$8.9 \times 10^{14}$~cm$^{-2}$.  This results in a $c$-C$_2$H$_3$DO to 
$c$-C$_2$H$_4$O ratio of $\sim$0.15 with the $c$-C$_2$H$_4$O column 
density taken from \citet{PILS_COMs_2017}. This corresponds to a 
D-to-H ratio of $\sim$0.036 per H atom because all four H atoms in 
$c$-C$_2$H$_4$O are equivalent with respect to substitution by one 
D. We constructed a rotation diagram from the unblended or only 
slightly blended transitions because these were numerous enough and 
span a range of upper state energies as can be seen in 
Table~\ref{tab_obs-lines}.  Such a rotation diagram may serve as a 
check of the modelling and it also provides a measure of the 
statistical uncertainties originating from the observational data. 
The rotation diagram is shown in Fig.~\ref{fig:rotdia} together with 
the results. We also applied our model based on the Band~7 data to the 
observations in Band~6: a small number of lines were detected in the 
more limited Band~6 data and only one appeared to be unblended 
(Fig.~\ref{band6_figure}). Still, the good match to the single 
transition is an additional verification of the results. These
transitions are also given in Table~\ref{tab_obs-lines}.


\begin{table}
  \begin{center}
  \caption{Quantum numbers, frequencies (MHz), upper state energies $E_{\rm up}$ (K) and Einstein 
           $A$ values ($10^{-4}$s$^{-1}$) of observed $c$-C$_2$H$_3$DO transitions$^a$ and notes.}
  \label{tab_obs-lines}
\smallskip
  \begin{tabular}{ccrccl}
  \hline
$J'_{K'_a,K'_c} - J''_{K''_a,K''_c}$ & Frequency  & $E_{\rm up}$ & $A$  & Notes \\
  \hline 
 $ 5_{5,1} -  4_{4,0}$               & 235180.99 &           33 & 1.95 &       \\
$22_{8,14} - 22_{7,15}$              & 235190.43 &          311 & 1.05 & $b,c$ \\
$22_{9,14} - 22_{8,15}$              & 235200.38 &          311 & 1.05 & $b,c$ \\
 $ 5_{5,0} -  4_{4,1}$               & 235965.82 &           33 & 1.95 & $b$   \\
 $ 9_{5,5} -  8_{4,4}$               & 331568.82 &           84 & 2.51 & $b,c$ \\
 $10_{5,5} -  9_{6,4}$               & 332130.07 &          103 & 1.26 &       \\
 $ 7_{7,1} -  6_{6,0}$               & 332520.25 &           63 & 5.99 &       \\
 $ 7_{7,0} -  6_{6,1}$               & 332594.25 &           63 & 6.00 &       \\
 $11_{2,9} - 10_{3,8}$               & 335611.61 &          105 & 5.09 &       \\
 $11_{3,9} - 10_{2,8}$               & 335620.97 &          105 & 5.09 &       \\
 $ 7_{4,3} -  6_{3,4}$               & 337191.43 &           54 & 0.79 & $c,d$ \\
 $10_{4,6} -  9_{5,5}$               & 338717.98 &          100 & 2.87 & $e$   \\
$12_{1,11} - 11_{2,10}$              & 345236.98 &          114 & 6.57 &       \\
$12_{2,11} - 11_{1,10}$              & 345237.01 &          114 & 6.57 &       \\
 $ 8_{6,3} -  7_{5,2}$               & 346866.07 &           72 & 3.76 &       \\
 $10_{5,6} -  9_{4,5}$               & 350271.77 &          100 & 3.35 &       \\
 $11_{3,8} - 10_{4,7}$               & 352688.86 &          112 & 4.94 &       \\
 $11_{4,8} - 10_{3,7}$               & 352939.91 &          112 & 4.95 &       \\
 $11_{6,5} - 10_{7,4}$               & 353520.70 &          125 & 0.95 & $b,c$ \\
$13_{0,13} - 12_{1,12}$              & 354922.73 &          122 & 8.10 &       \\
$13_{1,13} - 12_{0,12}$              & 354922.73 &          122 & 8.10 &       \\
 $ 8_{6,2} -  7_{5,3}$               & 358772.79 &           72 & 3.89 &       \\
$12_{2,10} - 11_{3,9}$               & 362232.22 &          122 & 6.65 &       \\
$12_{3,10} - 11_{2,9}$               & 362233.93 &          122 & 6.65 &       \\
  \hline 
\end{tabular}\\[2pt]
\end{center}
$^a$ Derived from our parameter set emplyoing the S reduction in the III$^l$ representation.\\
$^b$ Blended.\\
$^c$ Weak, close to the noise limit.\\
$^d$ Partially blended.\\
$^e$ Blended with absorption line.
\end{table}


\subsection{Discussion of astronomical results}
\label{astro-discussion}

The temperature of $103 \pm 19$~K determined from the rotation diagram $c$-C$_2$H$_3$DO 
agrees well with the 125~K obtained for $c$-C$_2$H$_4$O \citep{PILS_COMs_2017}. The column 
density from the rotation diagram of $(7.8 \pm 1.5) \times 10^{14}$~cm$^{-2}$ is slightly 
lower than $8.9 \times 10^{14}$~cm$^{-2}$ derived from the modelling applying 125~K 
to some degree because the partition function value is lower at 103~K
than at 125~K.

The degree of deuteration of $c$-C$_2$H$_3$DO referenced to one H atom of $\sim$0.036 is 
slightly larger than species such as formaldehyde, methanol, ketene, HNCO, formamide and 
cyanamide with ratios of 0.02--0.03 but lower than some of the larger complex species 
such as ethanol, methylformate, glycolaldehyde and acetaldehyde with ratios of 0.05--0.06 
\citep{PILS_div-isos_2018,PILS_formamide_2016,PILS_cyanamide_2018}. 
This difference may reflect differences in the formation time with the species with 
the lower ratios forming earlier in the evolution of the prestellar cores.

Some of the emission lines of monodeutero-oxirane are strong enough that it may be possible 
to detect doubly-deuterated oxirane. Recall that there are four different ways to substitute 
one H by one D, but all four of them are spectroscopically indistinguishable. 
There are six different ways to substitute two of the four H by two D, however, only two 
each are spectroscopically indistinguishable, leading to three spectroscopically different 
doubly-deuterated oxirane isotopomers, 1,1-dideutero-oxirane and $Z$-1,2-dideutero-oxirane 
with $C_{\rm S}$ symmetry and $E$-1,2-dideutero-oxirane having $C_2$ symmetry. 
If the degree of deuteration were also 0.036 per H atom in dideutero-oxirane this would lead 
to an abundance of 0.0027 relative to the main species for each of the three dideutero-oxirane 
isotopomers. Even the strongest lines would be close to the detection limit. 
Combined with a large probability that these lines are blended with stronger features of 
other species, the unambiguous detection of dideutero-oxirane in the PILS data would be 
fairly unlikely even if the presence of three isotopomers increase the chances to detect a 
sufficiently large number of lines with certainty. 
It is, however, common that the degree of deuteration per H atom is higher in doubly 
deuterated molecules compared to the respective singly deuterated isotopologues. 
In the case of D$_2$CO \citep{PILS_H2CO_2018}, CHD$_2$CN \citep{PILS_nitriles_2018} and 
CHD$_2$OCHO \citep{PILS_dideu-MeFo_2019} detected in the PILS data this degree is higher 
by factors of about 2.8, 3.3 and 2.0 respectively than for the corresponding singly deuterated 
isotopic species leading to column densities between a factor of four and ten higher than 
without this additional increase in deuteration. A similar result was obtained for the 
deuteration of NH$_2$ to ND$_2$ compared with that to NHD \citep{NHD_ND2_det_2020}. 
Additional support comes from the PILS data of CH$_3$OH and CH$_2$DOH 
\citep{PILS_overview_2016}, CHD$_2$OH \citep{CHD2OH_catalog_2022} and CD$_3$OH 
\citep{CD3OH_rot_2022}. It has been argued that the higher degree of deuteration 
per H atom for multiply deuterated isotopologues compared with the respective 
singly deuterated ones has been inherited from the prestellar phase. A very recent 
investigation into the deuteration of thioformaldehyde of the prestellar core 
L1544 \citep{H2CS-deuteration_L1544_2022} supports this view. 
If the deuteration of the three dideutero-oxirane isotopomers is enhanced in a similar 
way compared with that of monodeutero-oxirane we may observe column density ratios 
of 0.01 to 0.025 for each of the three dideutero-oxirane isotopomers compared to the 
main species, which may be enough for identifications provided not all of the 
stronger lines of these three species are blended with stronger lines of other species.


\section{Conclusions}
\label{conclusion}

We prepared a sample of $c$-C$_2$H$_3$DO to extend the line list and to improve the quality 
of its rest frequencies which is now sufficient well into the terahertz region.
We made use of the Protostellar Interferometric Line Survey of IRAS 16293$-$2422 to 
identify mono-deuterated oxirane towards its source~B at a level of $\sim$0.15 overall 
and $\sim$0.036 with respect to a single H atom relative to the main isotopologue, 
in good agreement with values observed for some organic molecules in this source 
but slightly higher than other molecules. 
The detection of the doubly deuterated oxirane isotopomers should be possible if 
the strongest lines are not all blended with stronger lines of other species.


\section*{Acknowledgements}

HSPM thanks Ziqiu Chen and Martin Quack for providing the $c$-C$_2$H$_3$DO millimetre 
data with higher accuracy. This paper makes use of the following ALMA data: 
ADS/JAO.ALMA$\#$2013.1.00278.S and ADS/JAO.ALMA\#2016.1.01150.S. ALMA is a partnership 
of ESO (representing its member states), NSF (USA) and NINS (Japan), together with NRC 
(Canada), NSC and ASIAA (Taiwan) and KASI (Republic of Korea), in cooperation with 
the Republic of Chile. The Joint ALMA Observatory is operated by ESO, AUI/NRAO and NAOJ. 
The work in K{\"o}ln was supported by the Deutsche Forschungsgemeinschaft through the 
collaborative research centre SFB~956 (project ID 184018867) project B3 and through 
the Ger{\"a}tezentrum SCHL~341/15-1 (``Cologne Center for Terahertz Spectroscopy''). 
The research of JKJ is supported by the European Research Council through the ERC 
Consolidator Grant ``S4F'' (grant agreement No~646908) and Centre for Star and Planet 
Formation funded by the Danish National Research Foundation.
J.-C. G. acknowledges support by the Centre National d'Etudes Spatiales 
(CNES; grant number 4500065585) and by the Programme National Physique et Chimie du 
Milieu Interstellaire (PCMI) of CNRS/INSU with INC/INP co-funded by CEA and CNES. 
Our research benefitted from NASA's Astrophysics Data System (ADS).


\section*{Data Availability}

The spectroscopic line lists and associated files are available as supplementary material 
through the journal and in the data section of the 
CDMS\footnote{https://cdms.astro.uni-koeln.de/classic/predictions/daten/Oxiran/}. 
The underlying original spectral recordings will be shared on reasonable request to 
the corresponding author. The radio astronomical data are available through the 
ALMA archive\footnote{https://almascience.eso.org/aq/}. 





\bibliographystyle{mnras}
\bibliography{hspm} 



\bsp	
\label{lastpage}
\end{document}